\documentclass[twocolumn,preprintnumbers,amsmath,amssymb,superscriptaddress]{revtex4}
\usepackage{graphicx}
\usepackage{dcolumn}
\usepackage{bm}
\usepackage{color}

\font\scripti=cmmi7
\font\scriptscripti=cmmi5
\def\sib#1{\setbox0 = \hbox{\scripti #1}
  \kern-.02em\copy0\kern-\wd0
  \kern.04em\box0} 
\def\ssib#1{\setbox0 = \hbox{\scriptscripti #1}
  \kern-.02em\copy0\kern-\wd0
  \kern.04em\box0} 
\font\tenib=cmmib10 
\skewchar\tenib='177 \skewchar\tenib='177 \skewchar\tenib='177
\def\pbold#1{\setbox0 = \hbox{$ #1 $}
  \kern-.022em\copy0\kern-\wd0
  \kern.011em\copy0\kern-\wd0
  \kern.011em\copy0\kern-\wd0
  \kern.011em\copy0\kern-\wd0
  \kern.011em\box0} 

\usepackage{graphicx}
\usepackage{dcolumn}
\usepackage{bm}

\def\up{\uparrow}
\def\dn{\downarrow}

\def\lesssim{\ \raise.3ex\hbox{$<$}\kern-0.8em\lower.7ex\hbox{$\sim$}\ }
\def\gesim{\ \raise.3ex\hbox{$>$}\kern-0.8em\lower.7ex\hbox{$\sim$}\ }

\begin{document}
\preprint{RIKEN-QHP-498}
\title{Unitary {\it p}-wave Fermi gas in one dimension}
\author{Hiroyuki Tajima}
\affiliation{Department of Physics, Graduate School of Science, The University of Tokyo, Tokyo 113-0033, Japan}
\author{Shoichiro Tsutsui}
\affiliation{RIKEN Nishina Center, Wako, Saitama, 351-0198, Japan}
\author{Takahiro M. Doi}
\affiliation{Research Center for Nuclear Physics (RCNP), Osaka University, Osaka 567-0047, Japan}
\author{Kei Iida}
\affiliation{Department of Mahematics and Physics, Kochi University, Kochi, 780-8520, Japan}

\date{\today}
\begin{abstract}
We elucidate universal many-body properties of a one-dimensional, two-component ultracold Fermi gas near the {\it p}-wave 
Feshbach resonance.  The low-energy scattering in this system can be characterized by two parameters, that is, {\it p}-wave 
scattering length and effective range.
At the unitarity limit where the {\it p}-wave scattering length diverges and the effective range is reduced to zero without conflicting with the causality bound, the system obeys universal thermodynamics as observed in a unitary Fermi gas with contact {\it s}-wave interaction in three dimensions.
It is in contrast to a Fermi gas with the {\it p}-wave resonance in three dimensions in which the effective range is inevitably finite.
We present the universal equation of state in this unitary {\it p}-wave Fermi gas within the many-body $T$-matrix approach as well as the virial expansion method.
Moreover, we examine the single-particle spectral function in the high-density regime where the virial expansion is no longer valid.
On the basis of the Hartree-like self-energy shift at the divergent scattering length,
we conjecture that {the equivalence of the Bertsch parameter across spatial dimensions} holds even for a one-dimensional unitary $p$-wave Fermi gas.


\end{abstract}
\maketitle
\par
\section{Introduction}
\label{sec1}
The concept of universality often facilitates approach to normally complicated many-body problems. 
While detailed structure of the interaction potential between constituent particles 
generally plays a crucial role in describing the properties of a specific many-body system, the possible 
presence of an infinitely large length scale such as the correlation length near a critical point and the scattering 
length near the Feshbach resonance could be a key to understanding the universality in  various systems.
For example, the Bardeen-Cooper-Schrieffer (BCS) to Bose-Einstein condensate (BEC) crossover~\cite{Eagles,Leggett,Nozieres,SadeMelo}, 
which is experimentally realized in ultracold atoms~\cite{Regal,Bartenstein,Zwierlein} (for recent review, 
see Refs.~\cite{Zwerger,Randeria,Strinati,Ohashi}), not only gives a unified way of understanding 
Fermi and Bose superfluidity in a specific system, but also an interdisciplinary viewpoint 
on how to deal with different systems with different energy scales such as strongly-correlated superconductors, superfluid helium,
nuclear matter, and color superconductors.
\par
Moreover, a unitary Fermi gas, which is located in the middle of the BCS-BEC crossover  where 
the $s$-wave scattering length diverges and hence the grand-canonical thermodynamic  potential 
includes only two energy scales, namely, the chemical potential $\mu$ and the temperature $T$, 
has been extensively investigated by theoretical~\cite{Ho} and experimental~\cite{Horikoshi,Nascimbene,Ku}
approaches.  In particular, the ground-state thermodynamic properties of such a gas are characterized 
by a single parameter called the Bertsch parameter~\cite{MBX,Baker}.
These universal ground-state properties have attracted attention from a viewpoint of the similarity 
between an ultracold Fermi gas and dilute neutron matter~\cite{Gandolfi,Carlson} 
and have been quantitatively revealed in cold atom experiments~\cite{Navon,HorikoshiX,HorikoshiE}.
At high temperatures, one can pin down the virial expansion coefficients~\cite{Liu}, 
which could be useful for describing matter in stellar collapse~\cite{Oertel}.
In this way, there is no doubt about the importance of further investigations on these universal states of matter.
Moreover, such a unitary Fermi gas is also predicted to occur in a one-dimensional system of 
four-component fermions with a four-body attraction~\cite{NishidaSon2010}.
Interestingly, it was reported from lattice simulations~\cite{Endres2012}
that the values of the Bertsch parameter in these unitary gases with different dimensions are 
close to each other.
In this context, other possibilities of realizing unitary gases with different kind of interactions such as $p$-wave resonant one are worth exploring.

While strongly interacting $s$-wave Fermi superfluids are successfully realized, the experimental realization of a superfluid Fermi gas near a $p$-wave Feshbach resonance~\cite{Regalp,Zhang,Gunter,Nakasuji} is still challenging.  In fact, $p$-wave superfluid Fermi gases, which have been anticipated over the past few decades~\cite{Ohaship,Gurarie}, suffer several difficulties such as strong three-body losses~\cite{Yoshida2018,Waseem2018,Waseem2019,Top,Schmidt}.  Fortunately, progress in experimental techniques has enabled {the observation of dipolar splittings~\cite{Ticknor,Gerken},} the formation of a $p$-wave molecule~\cite{Gaebler,Inada,Fuchs,Waseem2016}, 
and the measurement of the $p$-wave Tan's contact~\cite{Luciuk}.
Simultaneously, universal aspects of $p$-wave Fermi gases have attracted theoretical attention~\cite{Yu,Yoshida2015,Yoshida2016,Peng2016,Inotani}.
Also, it remains to be examined how atomic losses are suppressed in the presence of strong $p$-wave interaction.


To reach the unitarity limit, the scale of the effective range is important.
In a three-dimensional system, the effective range of the $p$-wave interaction is inevitably nonzero due to Wigner's causality bound~\cite{Wigner}.
Therefore, the unitarity limit can never be realized in a three-dimensional gas with a $p$-wave Feshbach resonance. 
In a one-dimensional system, the situation is totally different. 
Recently, the stabilization of $p$-wave Fermi gases confined in one spatial dimension~\cite{Zhou,Pan,Fonta} and in optical lattices~\cite{Han} has been pointed out theoretically,
while the three-body loss associated with a $p$-wave Feshbach resonance under low-dimensional confinement has experimentally been investigated~\cite{Chang2020,Marcum}.
The renormalization scheme of a contact-type $p$-wave interaction has also been presented in Refs.~\cite{Cui2016,Cui20162,Sekino2018,Sekino2020,Valiente1,Valiente2}.
Note that the low-energy scattering properties are generally classified by the combination $2L+d$ of the angular momentum $L$ and the dimension $d$~\cite{Hammer1,Hammer2}.
In this context, a one-dimensional system with a $p$-wave interaction ($L=1$ and $d=1$) belongs to the same class as a three-dimensional system 
with an $s$-wave interaction ($L=0$ and $d=3$).
Thus one can expect the realization of a unitary behavior in such a one-dimensional system as the unitarity limit has already been
achieved in the three-dimensional counterpart. 

In this work, we elucidate how the unitarity limit occurs for $L=1$ and $d=1$ and 
then examine the resulting universal many-body properties.
Physically, such unitarity would be realized {at zero temperature} because the interaction energy is of the same order as the kinetic energy with respect to the density.  In one dimension,
a typical momentum scale is given by the Fermi momentum $k_{\rm F}$, which is in turn proportional to the density.  Note that the size of a preformed pair
is given by the interparticle spacing rather than the scattering length in a many-particle system of interest here.  
According to the Lippmann-Schwinger equation, the effective interaction is then larger than the bare interaction by a factor of $1/(k_{\rm F} r_{\rm eff})$ with a vanishingly small effective range $r_{\rm eff}$.  
Since the bare interaction per fermion scales as $k_{\rm F}^3$, the effective interaction per fermion scales as $k_{\rm F}^2$, which is of the same order in $k_{\rm F}$
as the kinetic energy.


This paper is organized as follows.
In Sec.~\ref{sec2}, we present our formulation based on many-body $T$-matrix approach.
In Sec.~\ref{sec3}, we show the numerical results for the number density and single-particle spectral weight.
Finally, we conclude this paper in Sec.~\ref{sec4}.
Hereafter, we use units in which $\hbar=k_{\rm B}=1$ and the system volume is set to unity.


\section{Formulation}
\label{sec2}
We consider a one-dimensional two-component Fermi gas near a $p$-wave Feshbach resonance
and examine its equilibrium properties at chemical potential $\mu$ and temperature $T$.
The corresponding two-channel Hamiltonian reads
\begin{align}
\label{eq:H}
H&=\sum_{k,\sigma}\xi_{k}c_{k,\sigma}^\dag c_{k,\sigma}+\sum_{q}\xi_{q}^{\rm b}b_{q}^\dag b_{q}\cr
&+g\sum_{p,q}\left(pb_{q}^\dag c_{-p+q/2,\dn} c_{p+q/2,\up} + {\rm h. c.}\right),
\end{align}
where $\xi_k=k^2/(2m)-\mu$ and $c_{k,\sigma}^{(\dag)}$ are the kinetic energy minus the chemical potential and
and annihilation (creation) operator of a Fermi atom with mass $m$, momentum $k$, and pseudospin $\sigma=\up,\dn$, respectively.
{For simplicity, we consider an equal-mass mixture.}
The second term in the right-hand side of Eq.~(\ref{eq:H}) denotes the kinetic energy term of 
closed channel molecules with the energy level $\nu$, where $\xi_{q}^{\rm b}=q^2/(4m)-2\mu+\nu$ and 
$b_{q}^{(\dag)}$ are the kinetic energy minus the chemical potential and the annihilation (creation) operator of 
a bosonic molecule with momentum $q$.
The last term represents the $p$-wave (odd parity) Feshbach coupling with coupling constant $g$.
The $p$-wave scattering length $a$ and the effective range $r_{\rm eff}$ are related to $\nu$ and $g$ 
via the Lippmann-Schwinger equation as \cite{Cui2016,Cui20162}
\begin{align}
\frac{m}{2a}=-\frac{\nu_{\rm R}}{g^2}, \quad r_{\rm eff}=-\frac{4}{m^2g^2},
\label{eq:scatpara}
\end{align}
where $\nu_{\rm R}=\nu-g^2\frac{m\Lambda}{\pi}$ is the renormalized energy level of a closed channel molecule with the momentum cutoff $\Lambda$.
In this work we focus on the equally populated case.  In the presence of large spin polarization, Fermi polarons 
are expected to occur as has theoretically been investigated~\cite{Ma}.  In the perfectly polarized case,
an $s$-wave contact interaction does not work due to the Pauli principle, but a three-body interaction
is inevitable~\cite{Sekino2020}.  This is in contrast to the present case in which an $s$-wave interaction
is normally nonnegligible, but the three-body correlation is expected to be weak.  For a possible method for
eliminating the $s$-wave interaction, see Appendix in Ref.~\cite{Akagami}. 
{The intra-component $p$-wave interaction is not considered in this work. Such a situation is relevant for the $p$-wave resonance between different two hyperfine states~\cite{Gerken,Cui2016}.}

\par
We note that the present two-channel model, which involves the contact ($q$-independent) coupling $g$, reduces to the single-channel Hamiltonian with the effective $p$-wave interaction
$H_{\rm int.}=\sum_{p,p',q}pUp'c_{p+q/2,\up}^\dag c_{-p+q/2,\dn}^\dag c_{-p'+q/2,\dn} c_{p'+q/2,\up}$, 
where $U=-\frac{g^2}{\nu}$ approaches zero in the large $\Lambda$ limit.
This is in contrast to the $s$-wave case where the coupling constant is finite and 
therefore the Hartree shift gives a nonzero contribution~\cite{Tajima}.
\par
Many-body effects are incorporated into the self-energy $\Sigma_{\rm f}(p,i\omega_\ell)$ of a Fermi atom in the thermal Green's function $G(p,i\omega_\ell)=\left[\left\{G_0(p,i\omega_\ell)\right\}^{-1}-\Sigma_{\rm f}(p,i\omega_\ell)\right]^{-1}$, where $G_0(p,i\omega_\ell)=(i\omega_\ell-\xi_{p})^{-1}$ is the bare propagator with the fermion Matsubara frequency $\omega_\ell=(2\ell+1)\pi T$ ($\ell\in \mathbb{Z}$).
Within the many-body $T$-matrix approach,
$\Sigma_{\rm f}(p,i\omega_\ell)$ is given by
\begin{align}
\Sigma_{\rm f}(p,i\omega_\ell)&=T\sum_{q,i\nu_n}\Gamma\left(\frac{q}{2}-p,\frac{q}{2}-p;q,i\nu_n\right)\cr
&\times G_0(q-p,i\nu_n-i\omega_\ell),
\end{align}
where $\nu_n=2n\pi T$ is the boson Matsubara frequency ($n\in \mathbb{Z}$).
{The in-medium $T$-matrix 
\begin{align}
\Gamma(k,k';q,i\nu_n)=g^2kk'D(q,i\nu_n)
\end{align}
is associated with} 
the dressed molecular propagator
$D(q,i\nu_n)=\left[i\nu_n-\xi_{q}^{\rm b}-\Sigma_{\rm b}(q,i\nu_n)\right]^{-1}$.
The bosonic self-energy is given by $\Sigma_{\rm b}(q,i\nu_n)=g^2\Pi(q,i\nu_n)$, where
\begin{align}
\Pi(q,i\nu_n)=\sum_{p}p^2\frac{1-f(\xi_{p+q/2})-f(\xi_{-p+q/2})}{i\nu_n-\xi_{p+q/2}-\xi_{-p+q/2}}
\end{align} 
with the Fermi-Dirac distribution function $f(x)=(e^{x/T}+1)^{-1}$ is the lowest-order particle-particle bubble.
\par
In what follows,
we shall utilize the thermal momentum scale $k_{T}=\sqrt{2mT}$ associated with the temperature $T$ as well as the thermal de Broglie length $\lambda_T=\sqrt{\frac{2\pi}{mT}}\equiv 2\sqrt{\pi}k_T^{-1}$ for convenience.
Also, the dimensionless range parameter $R=|r_{\rm eff}k_T|$ will be used
to characterize the magnitude of $r_{\rm eff}$.
In this work, we will confine ourselves to the unitary system with $a^{-1}=0$.
\par
\section{Results}
\label{sec3}
\begin{figure}[t]
\begin{center}
\includegraphics[width=7cm]{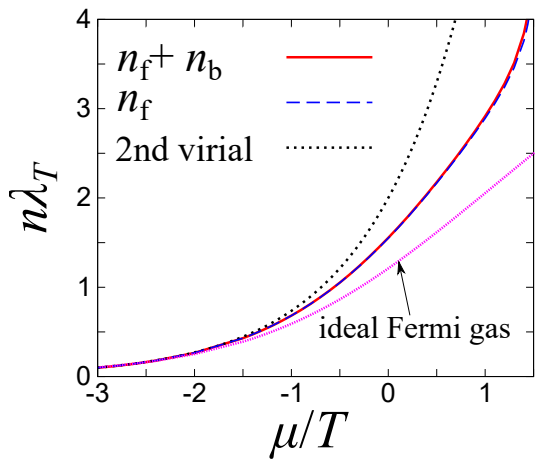}
\end{center}
\caption{
Number density equation of state for a unitary $p$-wave Fermi gas in one dimension.
The dotted curve (``2nd virial") shows the result from the second-order virial expansion.
{In this figure, we take $R=0.01$ which is sufficiently small to describe the universal regime.}
For reference, we also plot the behavior of an ideal Fermi gas.
}
\label{fig1}
\end{figure}
In the two-channel model, the number density $n$ reads $n=n_{\rm f}+n_{\rm b}$, where
$n_{\rm f}=2T\sum_{p,i\omega_\ell}G(p,i\omega_\ell)$ and $n_{\rm b}=-2T\sum_{q,i\nu_n}D(q,i\nu_n)$
are the fermionic and bosonic contributions, respectively.
Figure~\ref{fig1} shows the number density $n\lambda_T$ as a function of $\mu/T$ in the case of a negligibly small range parameter $R=0.01$.
As can be seen from Fig.~\ref{fig1}, $n_{\rm b}$ is negligibly small, which is natural because 
$n_{\rm b}$ reduces to zero in the limit of zero effective range ($R\rightarrow 0$).
The unitary gas in this limit strictly
obeys the universal thermodynamics in the sense that
the grand-canonical thermodynamic potential $\Omega(\mu,T)$ has no other energy scales than $\mu$ and $T$~\cite{Ho}.
In the low-density limit, $\Omega(\mu,T)$ can be obtained exactly by the virial expansion~\cite{Liu} as
$\Omega=-2\frac{T}{\lambda_T}\sum_{j=1}b_j z^j$, where $z=e^{\mu/T}$ is the fugacity. 
Within the second-order virial expansion, the number density reads
\begin{align}
n\lambda_T=2\left[b_1z+2b_2z^2+O(z^3)\right].
\end{align}
While the first order coefficient $b_1=1$ corresponds to the ideal classical gas contribution, the second one $b_2=b_2^{(0)}+\Delta b_2$ 
involves not only the non-interacting part $b_2^{(0)}=-\frac{1}{2\sqrt{2}}$ but also the interaction correction $\Delta b_2$.
$\Delta b_2$ can be obtained from the low-density limit of the many-body $T$-matrix approach, where $n$ is approximately given by
\begin{align}
\label{eq:n2}
n\simeq n_0 + 2T\sum_{p,i\omega_\ell}[G_0(p,i\omega_\ell)]^2\Sigma_{\rm f}(p,i\omega_\ell).
\end{align}
The second term in the right-hand side of Eq.~(\ref{eq:n2}) obtained by truncating the full Green's function up to 
first order in $\Sigma_{\rm f}(p,i\omega_\ell)$ is equivalent to the Nozi\`{e}res-Schmitt-Rink (NSR) correction $\delta n_{\rm NSR}$~\cite{Ohashi}.
Using the relation $\delta n_{\rm NSR}=\frac{4}{\lambda_T}\Delta b_2 z^2+O(z^3)$, we obtain $\Delta b_2=\frac{1}{2\sqrt{2}}$ and hence $b_2=0$ (see also Appendix~\ref{appA} for the derivation of $\Delta b_2$).
This indicates that the virial equation of state for a one-dimensional unitary $p$-wave Fermi gas up to second order in $z$ 
happens to be the same as the ideal classical one $\Omega=-2\frac{Tz}{\lambda_T}$ even in the presence of strong correlations. 
It is in contrast to a unitary Fermi gas in three dimensions where $b_2=\frac{3}{4\sqrt{2}}$~\cite{Liu}.
We note that the number density of an ideal Fermi gas is smaller than the unitary gas result due to the lack of pairing fluctuations. 
\par
While the virial expansion is no longer valid for $z>1$ ($\mu>0$), corresponding to a quantum degenerate regime of the unitary Fermi gas,
the many-body $T$-matrix approach in this regime is still expected to give a semi-quantitative description of strong coupling effect such as emergence of a pairing pseudogap. 
\begin{figure}[t]
\begin{center}
\includegraphics[width=7cm]{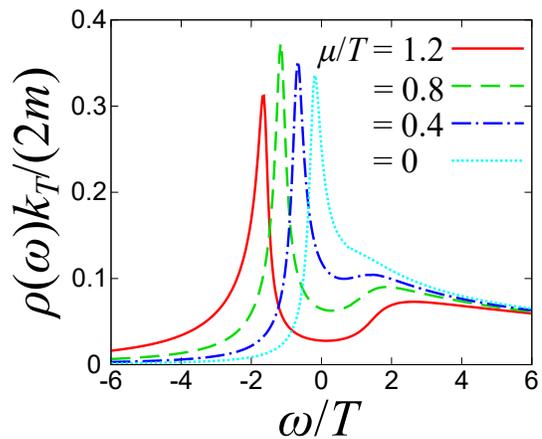}
\end{center}
\caption{Single-particle density of states $\rho(\omega)$ as a function of the single-particle energy $\omega$ in a unitary $p$-wave Fermi gas where $a^{-1}=0$.}
\label{fig2}
\end{figure}
In Fig.~\ref{fig2} we present the single-particle density of states $\rho(\omega)=\sum_{k}A(k,\omega)$ in such a unitary $p$-wave Fermi gas within the many-body $T$-matrix approach, where the single-particle spectral weight $A(k,\omega)=-\frac{1}{\pi}{\rm Im}G(k,i\omega_\ell\rightarrow \omega+i\delta)$ is obtained by the analytic continuation of $G(k,i\omega_\ell)$ to the real frequency $\omega$ ($\delta$ is a positive infinitesimal).
At zero chemical potential, $A(k,\omega)$ exhibits a single-particle peak near $\omega=0$, which is a specific behavior in one dimension.
At larger $\mu$, $\rho(\omega)$ shows the pseudogap opening around $\omega=0$ due to strong pairing fluctuations.
Although emergence of a pseudogap is still under debate in a three-dimensional unitary Fermi gas~\cite{Jensen,Halford}, the present system is expected to have a pseudogap pairing enhanced by low-dimensional fluctuations~\cite{Tajima}.
We note that a possible pseudogap in a three-dimensional Fermi gas with $p$-wave interaction has also been discussed in Ref.~\cite{Inotani2012}. 

\begin{figure}[t]
\begin{center}
\includegraphics[width=8cm]{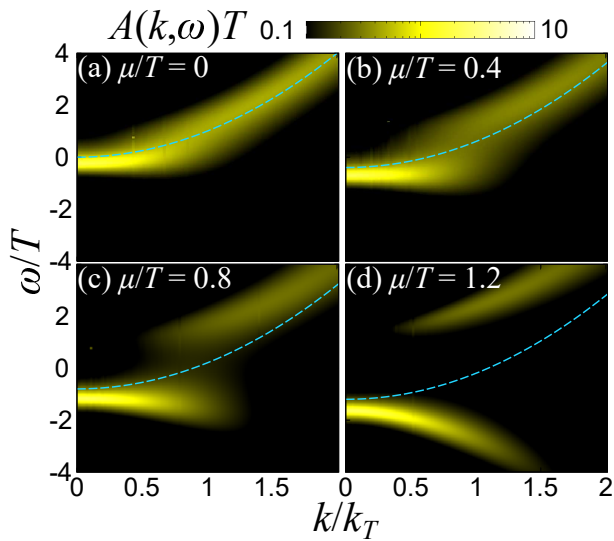}
\end{center}
\caption{Single-particle spectral weight $A(k,\omega)$ in a unitary $p$-wave Fermi gas in one dimension.
The dashed curve represents the single-particle dispersion $\omega=k^2/(2m)-\mu$ in an ideal Fermi gas. 
}
\label{fig3}
\end{figure}
More detailed single-particle excitations can be found in $A(k,\omega)$, which is shown in Fig.~\ref{fig3}.
One can see that the single-particle branch in the low-density regime ($\mu/T=0$) is separated into two branches due to strong pairing fluctuations, leading to the pseudogap opening in $\rho(\omega)$ as shown in Fig.~\ref{fig2}.
A similar spectral structure can be found in the case of the $s$-wave interaction in one dimension~\cite{Tajima}.
While the peak position in $A(k,\omega)$ at $\mu/T=0$ is close to the non-interacting dispersion $\omega=\xi_k$, 
at sufficiently large chemical potential where a pseudogap appears, deviation of the two branchs from the non-interacting dispersion increases with $\mu/T$.

\begin{figure}[t]
\begin{center}
\includegraphics[width=8cm]{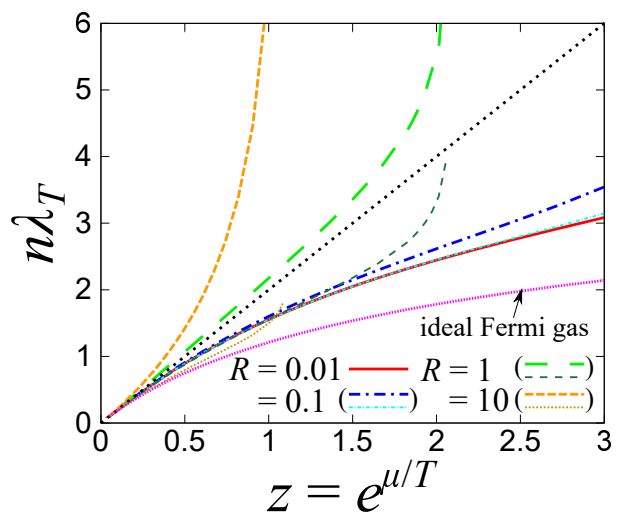}
\end{center}
\caption{
Non-universal effect on the number density $n\lambda_T$, which can be seen by
plotting $n=n_{\rm f}+n_{\rm b}$ (thick curves) and $n_{\rm f}$ (thin curves) at $R=0.1$, $1$, and $10$. 
For the result at $R=0.01$ we do not show $n_{\rm f}$ since $n_{\rm b}$ is negligibly small.
{The calculations are stopped at $\Delta_{\rm TC}=0$ (see Eq.~(\ref{eqdtc}) and the following text), indicating that the $T$-matrix approach breaks down.}
The diagonal dotted curve represents the second-order virial result at $R=0$, which
is the same as the behavior of an ideal classical gas, while the behavior of an ideal Fermi gas
is also plotted for comparison.
}
\label{fig4}
\end{figure}
We now turn to the non-universal effect of nonzero effective range on thermodynamic quantities.
Figure~\ref{fig4} shows how $n\lambda_T$ behaves with increasing $R$ as a function of $z$.
For $R\gtrsim0.1$, the fraction of molecules in the closed channel, $n_{\rm b}/n$, is no longer negligible.
Since $\Sigma_{\rm b}(q,i\nu_n)$ is proportional to $R^{-1}$, one can qualitatively estimate $n_{\rm b}\approx 2\sum_{q}e^{-\left\{\frac{q^2}{4m}-2\mu+\nu-\Sigma_{\rm b}(0,0)\right\}/T} \propto z^2e^{-\frac{\alpha}{R}}$ (with the constant $\alpha>0$), indicating that $n_{\rm b}$ becomes exponentially large when $R$ increases. 
On the other hand, one can observe that $n_{\rm f}$ with finite $R$ is close to the zero effective range result {(corresponding to the solid curve with $R=0.01$ in Fig.~\ref{fig4})},
indicating that the open-channel fraction is essential to extract the universal part of the equation of state for this system.
\par
We mention the limitation of the present many-body $T$-matrix approach to this one-dimensional system.
In fact, the numerical results for the number density in the high-density regime in Fig.~\ref{fig4} are difficult to obtain by this limitation. 
The many-body $T$-matrix approach breaks down when the dimensionless indicator of the Thouless criterion~\cite{Thouless} defined by
\begin{align}
\label{eqdtc}
\Delta_{\rm TC}= \frac{kk'}{\Gamma(k,k,q=0,i\nu_n=0)}\frac{\pi}{mk_T}
\end{align}
becomes zero.
Although in a three-dimensional system, an infrared divergence of the $T$-matrix
would indicate the occurrence of a superfluid phase transition,
this divergence in one dimension is an artifact of theory because such a phase transition {in one-dimensional systems} is prohibited by the Mermin-Wagner-Hohenberg theorem~\cite{MW,Hohenberg}.
Thus, $\Delta_{\rm TC}=0$ gives {an artificial} critical value of $\mu/T$ above which the present approach is no longer available.
\begin{figure}[t]
\begin{center}
\includegraphics[width=7cm]{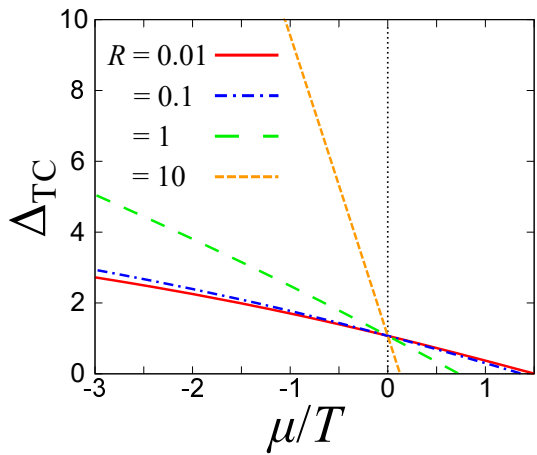}
\end{center}
\caption{Dimensionless indicator $\Delta_{\rm TC}$ for the Thouless criterion at $R=0.01$, $0.1$, $1$, and $10$. In this figure, we take $a^{-1}=0$. }
\label{fig5}
\end{figure}
Figure~\ref{fig5} shows $\Delta_{\rm TC}$ as a function of $\mu/T$ at $R=0.01$, $0.1$, $1$, and $10$.
In the unitarity limit where $R$ is negligible, corresponding to the result at $R=0.01$ in Fig.~\ref{fig5}, $\Delta_{\rm TC}$ becomes zero around $\mu/T=1.5$.
Since the correction due to the effective range, which is negative as shown in Eq.\ (\ref{eq:scatpara}), induces strong pairing in the two-channel model~\cite{Ho2012,Tajima2,Tajima3} and hence strong thermal fluctuations, however,
this critical $\mu/T$ itself is doubtful.
To safely avoid the above-mentioned artifact in the case of small $R$, therefore, one has to consider higher-order fluctuations beyond the present approach, which are left for future work.
\par
In the large $R$ limit, on the other hand, $2\mu-\nu$ cannot be positive.
In this limit, which is equivalent to $g\rightarrow 0$, the number density is exactly given by
\begin{align}
\label{eq:nlargeR}
n=2\sum_{p}f(\xi_{p})+2\sum_{q}b(\xi_q^{\rm b}),
\end{align}
where $b(x)=(e^{x/T}-1)^{-1}$ is the Bose-Einstein distribution function.
Obviously, $\xi_{q}^{\rm b}$ cannot be negative for any $q$ due to the Bose statistics in the second term of the right-hand side of Eq.~(\ref{eq:nlargeR}), resulting 
in $\mu <\nu/2$ {such that $\exp{\left(\frac{q^2}{4mT}+\frac{\nu-2\mu}{T}\right)}-1> 0$ for an arbitrary $q$}. 
This gives an exact upper bound for $\mu$ in the limit of $R\rightarrow \infty$, in contrast to the case of $R\rightarrow 0$
in which the condition $\Delta_{\rm TC}=0$ originates from the artifact of the present theoretical approach.
{Thus, the exact upper bound for $\mu$ at $|a|\rightarrow \infty$ is expected to be a function of $R$ which requires $\mu<\nu/2=0$ at $R\rightarrow \infty$  and gives no constraints on $\mu$ at $R\rightarrow 0$. We note that the critical $\mu$ in the $T$-matrix approach identified by $\Delta_{\rm TC}=0$ is smaller than this exact upper bound at finite~$R$. }

{
In spite of such limitations of the present approach, the Bertsch parameter might be considered by
utilizing the knowledge of a three-dimensional unitary Fermi gas.  In three dimensions,
one can estimate the ground-state chemical potential {divided} by the Fermi energy $E_{\rm F}$, that is, the Bertsch parameter $\xi_{\rm B}^{\rm 3DS}$ by replacing the $s$-wave scattering length $a_s$ with $k_{\rm F}^{-1}$ in the Hartree shift~\cite{Heiselberg,Kinnunen}.
Indeed, one can estimate $\xi_{\rm B}^{\rm 3DS}\simeq\frac{4\pi k_{\rm F}^{-1}}{mE_{\rm F}}\frac{n}{2}\simeq 0.424$, which is {fairly} close to the experimental values $\sim 0.4$~\cite{Ku,Navon,HorikoshiX,Nascimbene}.
This implies that the characteristic length scale for the interaction is approximately given by the interparticle distance being proportional to $k_{\rm F}^{-1}$ at $|a|\rightarrow \infty$.
In the present one-dimensional $p$-wave unitary gas,
the Hartree self-energy $\Sigma_{\rm H}(p,a)$ is given by
\begin{align}
\Sigma_{\rm H}(p,a)&=\frac{2a}{m}\sum_{q}\left(\frac{q}{2}-p\right)^2f(\xi_{q-p})\cr
&=\frac{ak_{\rm F}^3}{6\pi m} +\frac{ak_{\rm F}}{2\pi m}p^2
\ \ (T\rightarrow 0).
\end{align}
{By following the same line of argument of the} three-dimensional case, we obtain the corresponding Bertsch parameter $\xi_{\rm B}^{\rm 1DP}$ as
\begin{align}
\xi_{\rm B}^{\rm 1DP}&\simeq
\frac{\Sigma_{\rm H}(p=k_{\rm F},a=k_{\rm F}^{-1})}{E_{\rm F}}\cr
&\simeq 0.424,
\end{align}
where we {have} assumed that the self-energy shift at $p=k_{\rm F}$ is relevant for our purpose.
Surprisingly, it is completely equal to {the value of} $\xi_{\rm B}^{\rm 3DS}$ based on the same ansatz {as} mentioned above.
This result suggests that the transdimensional equivalence of the Bertsch parameter~\cite{Endres2012} might apply even to the one-dimensional $p$-wave case.
It is interesting to check the validity of this conjecture {and examine deviation of the predicted $\xi_{\rm B}^{\rm 1DP}$ from the exact value}, which will be addressed elsewhere.  { Also, it is useful to note that
experiments} for realizing the present unitary $p$-wave Fermi gas {are} expected to be {more feasible than the case of} a four-component unitary Fermi gas with a four-body attraction in one dimension~\cite{NishidaSon2010,Endres2012}.
}
\section{Conclusion}
\label{sec4}
We have theoretically investigated universal many-body properties of a one-dimensional two-component unitary Fermi gas with a $p$-wave contact interaction.
Thermodynamic functions in this unitary gas exhibit the universal behavior as in the case of a three-dimensional unitary Fermi gas with an $s$-wave short-range interaction. 
We have obtained the universal equation of state in the limit of zero effective range within the many-body $T$-matrix approach and derived the exact result for the second-order virial expansion, which interestingly is equivalent to the ideal classical gas result.
Even in the case of finite effective range, the number density of open-channel fermions is close to the universal result.
Moreover, we have shown that strong pairing fluctuations are visible as the pseudogap opening in a single-particle spectral weight.
{Finally, on the basis of the Hartree-like energy shift at unitarity, we have conjectured that the transdimensional equivalence of the Bertsch parameter holds even for this exotic unitary Fermi gas.
}

\par
For future theoretical perspective, it is interesting to address the Bertsch parameter in this $p$-wave unitary Fermi gas in more sophisticated manner, which could be addressed by lattice simulations, {the} thermodynamic Bethe {ansatz}, variational approaches, as well as future experiments.
{It is worth investigating a bosonic counterpart via the Bose-Fermi mapping~\cite{Girardeau}.}
Indeed, a similar spin-$1/2$ system with intra-component $p$-wave interactions has already been studied by using the Bose-Fermi mapping~\cite{Girardeau2006,Campo}.
Moreover, a similar unitarity limit {is expected to occur} in a spin-polarized {one-dimensional} Fermi gas with $p$-wave {contact} interaction.
We note, {however,} that in such a case the necessity of a three-body force for the renormalization has been pointed out in Ref.~\cite{Sekino2020}.
In this regard, it {would be essential to investigate} how three-body correlations occur {simultaneously.}
To achieve a $p$-wave unitary Fermi gas experimentally, {on the other hand,} the suppression of {other} residual interactions {would be} important. 
\par
{Apart from cold atomic physics, it is also interesting to consider the applications to other one-dimensional systems with $p$-wave interactions such as confined $^3$He fluids~\cite{Yager,Matsushita}
and unconventional superconductors~\cite{Pang}.
}

 
\acknowledgements
The authors thank Yuta Sekino for useful discussions.
This work was supported by Grant-in-Aids for Scientific Research provided by JSPS through Nos.~18H05406 and 18H01211 and for Early-Career Scientists through No.~20K14480. 
S. T. was supported by the RIKEN Special Postdoctoral Researchers Program. 
\par

\appendix
{
\section{Second-order virial coefficient at $p$-wave unitarity}
\label{appA}
In this appendix, we present a detailed derivation of the second-order virial coefficient $b_2$ {in the limit of zero effective range} from the many-body $T$-matrix approach.
First, the non-interacting contribution $b_2^{(0)}$ can be obtained by expanding $n_0$ with respect to $z$ as
\begin{align}
\label{eq:a1}
n_0&=2\sum_{p}f(\xi_p)\cr
&=2\frac{z}{\lambda_T}-4\frac{1}{2\sqrt{2}}\frac{z^2}{\lambda_T}+O(z^3).
\end{align}
From Eq.~(\ref{eq:a1}), 
one can find $b_2^{(0)}=-\frac{1}{2\sqrt{2}}$.
The second term in {the right-hand side of} Eq.~(\ref{eq:n2}) can be rewritten as
\begin{align}
\label{eq:a2}
\delta n=-T\sum_{q,i\nu_s}\gamma(q,i\nu_s)\frac{\partial}{\partial \mu}\gamma^{-1}(q,i\nu_s),
\end{align}
where $\gamma(q,i\nu_s)=\Gamma(k,k'q,i\nu_s)/(kk')$.
The summation with respect to $i\nu_s$ can be replaced with the {integral along the contour $C$} enclosing the imaginary energy axis as
\begin{align}
\label{eq:a3}
\delta n&=-\sum_{q}\oint_C\frac{d\zeta b(\zeta)}{2\pi i}\gamma(q,\zeta)\frac{\partial}{\partial \mu}\gamma^{-1}(q,\zeta)\cr
&=-\sum_{q}\oint_{C'}\frac{d\zeta' b(\zeta'+\epsilon_{q}^{\rm b})}{2\pi i}\bar{\gamma}(q,\zeta')\frac{\partial}{\partial \mu}\bar{\gamma}^{-1}(q,\zeta'),
\end{align}
where we {have} changed the variable $\zeta\rightarrow \zeta'+\epsilon_{q}^{\rm b}$ ($\epsilon_{q}^{\rm b}=\frac{q^2}{4m}-2\mu$) and the contour $C\rightarrow C'$ which encloses the pole $\zeta'=i\nu_s-\epsilon_{q}^{\rm b}$.
Also, we {have} used $\bar{\gamma}(q,\zeta')=\gamma(q,\zeta'+\epsilon_{q}^{\rm b})$.
Since we are interested in the correction of $O(z^2)$ and the lowest order of $b(\zeta'+\epsilon_q^{\rm b})\simeq z^2e^{-\frac{\zeta'}{T}}e^{-\frac{q^2}{4mT}}$ is already $O(z^2)$,
we can safely neglect the medium correction in $\bar{\gamma}(q,\zeta')$, leading to
\begin{align}
\label{eq:a4}
\bar{\gamma}(q,\zeta')\simeq \left[\frac{m}{2a}+i\frac{m}{2}\sqrt{m\zeta'}\right]^{-1}
\end{align}
and $\frac{\partial \bar{\gamma}^{-1}(q,\zeta')}{\partial \mu}=2\frac{\partial \bar{\gamma}^{-1}(q,\zeta')}{\partial \zeta'}=\frac{m^2}{2\sqrt{m\zeta'}}i$.
Using {these relations} at unitarity $1/a=0$,
we can obtain
\begin{align}
\label{eq:a5}
\delta n&=-z^2\sum_{q}e^{-\frac{q^2}{4mT}}\oint_{C'}\frac{d\zeta'}{2\pi i}\frac{e^{-\zeta'/T}}{\zeta'}+O(z^3),\cr
&=4\frac{1}{2\sqrt{2}}\frac{z^2}{\lambda_T}+O(z^3),
\end{align}
which {leads to} $\Delta b_2=\frac{1}{2\sqrt{2}}$.
Finally, combining these results, one can obtain the vanishing second-order coefficient $b_2=b_2^{(0)}+\Delta b_2=0$.
}

\end{document}